\documentclass[smallextended]{svjour3}       
\smartqed  
\usepackage{amsfonts}

\begin{document}

\title{Quasi-local energy from a Minkowski reference}

\author{Chiang-Mei Chen \and Jian-Liang Liu \and James M. Nester}

\institute{C.-M. Chen \at
              Department of Physics, National Central University, Chungli 32001, Taiwan \\
              Center for High Energy and High Field Physics (CHiP), National Central University, Chungli 32001, Taiwan \\
              \email{cmchen@phy.ncu.edu.tw}
           \and
           J.-L. Liu \at
              Department of Mathematics and Data Science, Dongguan University of Technology, Dongguan, China \\
              Department of Mathematics, Shantou University, Shantou, Guandong, 515063, China \\
              \email{2018013@dgut.edu.cn}
           \and
           J. M. Nester \at
              Department of Physics, National Central University, Chungli 32001, Taiwan \\
              Graduate Institute of Astronomy, National Central University, Chungli 32001, Taiwan \\
              Leung Center for Cosmology and Particle Astrophysics, National Taiwan University, Taipei 10617, Taiwan \\
              \email{nester@phy.ncu.edu.tw}
}

\date{Received: date / Accepted: date}

\maketitle

\begin{abstract}
The specification of energy for gravitating systems has been an unsettled issue since Einstein proposed his pseudotensor. It is now understood that energy-momentum is \emph{quasi-local} (associated with a closed 2-surface). Here we consider quasi-local proposals (including pseudotensors) in the Lagrangian-Noether-Hamiltonian formulations. There are two ambiguities: (i) there are many possible expressions, (ii) they depend on some non-dynamical structure, e.g., a reference frame. The Hamiltonian approach gives a handle on both problems.
The Hamiltonian perspective helped us to make a remarkable discovery:
with an isometric Minkowski reference a large class of expressions---namely all those that agree with the Einstein pseudotensor's Freud superpotential to linear order---give a common quasi-local energy value. Moreover, with a best-matched reference on the boundary this is the Wang-Yau mass value.

\end{abstract}

\section{Introduction}

Along with his celebrated general relativity (GR) field equations, in November 1915 Einstein also proposed his gravitational energy-momentum density, which is a \emph{pseudotensor}, not a proper tensor with coordinate reference frame independent meaning~\cite{cpae}.
Many objected, including Lorentz, Levi-Civita, Felix Klein and Schr\"odinger (he later called it ``sham''~\cite{Schroedinger}).
Einstein understood their concerns, but believed that his pseudotensor had physical meaning.
Emmy Noether's paper with her two famous theorems concerning symmetry in dynamical systems was written to clarify issues regarding energy raised by the investigations of Einstein, Hilbert and Klein~\cite{Rowe1999,Kosmann-Schwarzbach}.
She conclusively showed that there is no proper conserved energy-momentum density for any geometric gravity theory.

The topic of gravitational energy has remained an outstanding puzzle for over a century.
Various pseudotensor and quasi-local expressions obtained from different perspectives have been proposed.
Those that fit into the Lagrangian-Noether-Hamiltonian framework all depend on some non-dynamic structure, e.g., a reference frame.
Here we present in detail our recent discovery, which was briefly described in the essay~\cite{Chen:2018geu},  regarding how choosing an isometric Minkowski reference sheds new light on the gravitational energy issue.  In order to place our discovery in a suitable framework where it can be easily verified and its import can be better appreciated, we believe it appropriate to first review some of the history, survey many of the proposed energy expressions, explain how the Hamiltonian perspective clarifies the issues, and briefly describe our covariant Hamiltonian approach.

\section{The pseudotensors}

The Einstein Lagrangian density is quadratic in the connection, it differs from Hilbert's scalar curvature Lagrangian by a total divergence:\footnote{Our notation follows in general MTW~\cite{MTW}.}
\begin{equation} \label{EinsteinL}
2 \kappa{\cal L}_{\rm E} := - \sqrt{|g|} g^{\beta\sigma} \Gamma^\alpha{}_{\gamma\mu} \Gamma^\gamma{}_{\beta\nu} \delta^{\mu\nu}_{\alpha\sigma} \equiv \sqrt{|g|} R - \hbox{div}
\end{equation}
(here $\kappa := 8 \pi G/c^3$).
It depends on the metric and its first derivatives.
The \emph{Einstein pseudotensor} is the associated canonical energy-momentum density:
\begin{equation}
\mathfrak{t}_{\rm E}^\mu{}_\nu := \delta^\mu_\nu{\cal L}_{\rm E} - \frac{\partial{\cal L}_{\rm E}}{\partial \partial_\mu g_{\alpha\beta}} \partial_\nu g_{\alpha\beta},
\qquad
\partial_\mu\mathfrak{t}_{\rm E}^\mu{}_\nu \equiv \frac{\delta{\cal L}_{\rm E}} {\delta g_{\alpha\beta}} \partial_\nu g_{\alpha\beta}, \label{t}
\end{equation}
where $\delta{\cal L}_{\rm E}/\delta g_{\alpha\beta}=0$ is the Einstein equation.  With a source it reads
\begin{equation}
\sqrt{|g|} G^\mu{}_\nu = \kappa \mathfrak{T}^\mu{}_\nu := \kappa \sqrt{|g|} T^\mu{}_\nu.
\end{equation}
Applying the contracted Bianchi identity yields
$\nabla_\mu \mathfrak{T}^\mu{}_\nu = 0$. With~(\ref{t}) this gives a (noncovariantly) conserved total energy-momentum complex:
\begin{equation}
\partial_\mu (\mathfrak{T}^\mu{}_\nu + \mathfrak{t}_{\rm E}^\mu{}_\nu) = 0.
\end{equation}
From this relation one can infer the existence of some \emph{superpotential} $\mathfrak{U}^{\mu\lambda}{}_\nu \equiv \mathfrak{U}^{[\mu\lambda]}{}_\nu$ such that
\begin{equation}
\kappa^{-1} \sqrt{|g|} G^\mu{}_\nu + \mathfrak{t}_{\rm E}^\mu{}_\nu = \partial_\lambda \mathfrak{U}^{\mu\lambda}{}_\nu.\label{superpot}
\end{equation}
Such a superpotential was presented only in 1939 by Freud~\cite{Freud,Boehmer:2017tvw}:
\begin{equation} \label{UF}
2 \kappa \mathfrak{U}_{\rm F}^{\mu\lambda}{}_\nu := - \sqrt{|g|} {g}^{\beta\sigma}
\Gamma^\alpha{}_{\beta\gamma} \delta^{\mu\lambda\gamma}_{\alpha\sigma\nu} \equiv - \sqrt{|g|} g^{\beta\sigma} g^{\alpha\delta} \delta^{\mu\lambda\gamma}_{\alpha\sigma\nu} \partial_\beta g_{\delta\gamma}.
\end{equation}
Other pseudotensors were proposed from various perspectives by Landau-Lifshitz~\cite{LL}, Papapetrou~\cite{Papapetrou:1948jw}, Bergmann-Thomson~\cite{Bergmann:1953jz}, Goldberg~\cite{Goldberg:1958zz}, M{\o}ller~\cite{Mol58}, and Weinberg~\cite{Wein72} (also used in~\cite{MTW}). They likewise follow from their respective superpotentials. Here, with the aid of the Minkowski reference metric $\bar g_{\mu\nu}$ and its associated covariant derivative $\bar\nabla_\mu$, we list the well-known superpotentials; all are written as weight one tensor densities\footnote{These superpotentials can all be put into the standard form of $2\choose1$ type expressions by using the Minkowski metric $\bar g_{\alpha\beta}$ to lower the $\nu$ index.}
in forms that reveal their interrelationships:
\begin{eqnarray}
2 \kappa \mathfrak{U}_{\rm F}^{\mu\lambda}{}_\nu &:=& -\sqrt{|g|} {g}^{\beta\sigma} (\Gamma^\alpha{}_{\beta\gamma} - \bar\Gamma^\alpha{}_{\beta\gamma}) \delta^{\mu\lambda\gamma}_{\alpha\sigma\nu}
\equiv - \sqrt{|g|} g^{\beta\sigma} g^{\alpha\delta} \delta^{\mu\lambda\gamma}_{\alpha\sigma\nu} {\bar\nabla}_\beta g_{\delta\gamma}, \label{UF3}
\\
2 \kappa \mathfrak{U}_{\rm LL}^{\mu\lambda\nu} &:=& |\bar g|^{-\frac12} \delta^{\mu\lambda}_{\gamma\alpha} \bar\nabla_\pi[|g| g{}^{\alpha\nu} g^{\gamma\pi}] \equiv \delta^{\mu\lambda}_{\gamma\alpha} \delta^{\nu\pi}_{\beta\rho} |g/\bar g|^{\frac12} g^{\alpha\beta} \bar\nabla_\pi(|g|^{\frac12} g^{\gamma\rho}), \label{LL}
\\
2 \kappa \mathfrak{U}_{\rm P}^{\mu\lambda\nu} &:=& \delta^{\mu\lambda}_{\gamma\alpha} \delta^{\nu\pi}_{\beta\rho} \bar g{}^{\alpha\beta} \bar\nabla_\pi[|g|^{\frac12} g^{\gamma\rho}]
\nonumber\\
&& \equiv \delta^{\mu\lambda}_{\gamma\alpha} \delta^{\nu\pi}_{\beta\rho} \bar g{}^{\alpha\beta} |g|^{\frac12} ({\textstyle\frac12} g^{\gamma\rho} g^{\tau\delta} - g^{\gamma\tau} g^{\rho\delta}) {\bar\nabla}_\pi g_{\tau\delta}, \label{HP}
\\
\mathfrak{U}_{\rm BT}^{\mu\lambda\nu} &:=& g^{\nu\delta} \frak{U}_{\rm F}^{\mu\lambda}{}_\delta \equiv |\bar g/g|^{\frac12} \frak{U}_{\rm LL}^{\mu\lambda\nu}, \label{UBT}
\\
\mathfrak{U}_{{\rm F}_n}^{\mu\lambda}{}_\nu &:=& |g/\bar g|^{\frac{n}2} \mathfrak{U}_{\rm F}^{\mu\lambda}{}_\nu, \label{UFn}
\\
\mathfrak{U}_{{\rm LL}_n}^{\mu\lambda\nu} &:=& |\bar g|^{-\frac12} \delta^{\mu\lambda}_{\gamma\alpha} \bar\nabla_\pi[|g||g/\bar g|^{\frac{n}2} g{}^{\alpha\nu} g^{\gamma\pi}]
\nonumber\\
&& \equiv |g/\bar g|^{\frac{n}2} \mathfrak{U}_{\rm LL}^{\mu\lambda\nu} +
|g|^{\frac12} |g/\bar g|^{\frac{n+1}2} \delta^{\mu\lambda}_{\gamma\alpha} g^{\alpha\nu} g^{\gamma\pi} g^{\sigma\delta}
\bar\nabla_\pi g_{\sigma\delta}, \label{ULLn}
\\
2 \kappa \mathfrak{U}_{\rm M}^{\mu\lambda}{}_\nu &:=& 2|g|^{\frac12} \delta^{\mu\lambda}_{\alpha\sigma} g^{\beta\alpha} g^{\sigma\delta} {\bar\nabla}_\beta g_{\delta\nu} \label{UM},
\\
2 \kappa \mathfrak{U}_{\rm W}^{\mu\lambda\nu} &:=& \delta^{\mu\lambda}_{\gamma\alpha} \delta^{\nu\pi}_{\beta\rho} \bar g{}^{\alpha\beta} |\bar g|^{\frac12} ({\textstyle\frac12} {\bar g}^{\gamma\rho} {\bar g}^{\tau\delta} - {\bar g}^{\gamma\tau} {\bar g}^{\rho\delta}) {\bar\nabla}_\pi g_{\tau\delta}. \label{HW}
\end{eqnarray}
This list includes Goldberg's Freud~(\ref{UFn}) and Landau-Lifshitz~(\ref{ULLn}) density-weighted superpotentials.
In a Minkowski coordinate frame, $y^{\bar\mu}$, where $\bar g_{\bar\mu\bar\nu} = {\rm diag}(-1,+1,+1,+1)$, $\bar\nabla_{\bar\mu} = \partial_{\bar\mu}$, $\bar\Gamma^{\bar\alpha}{}_{\bar\beta\bar\gamma} = 0$ and $|\bar g| = 1$, the above superpotential expressions revert to their traditional form.
All of these superpotentials define energy-momentum values which depend on the Minkowski reference geometry, or equivalently a Minkowski coordinate system.

There are two unsatisfactory issues: (i) which expression? (ii) and which Minkowski reference geometry/coordinate system?
On the other hand (a) these expressions do provide a description of energy-momentum conservation, (b) they (like connections) have well defined values for each Minkowski reference geometry/Minkowski coordinate reference frame, (c) all (except for M{\o}ller) give the expected total energy-momentum values at spatial
infinity.\footnote{Ref.~\cite{MTW}, \S 20.2 uses the linearized theory to argue that the total energy-momentum for an asymptotically flat spacetime is given by the integral over the 2-surface boundary at infinity of effectively the expression (\ref{HW}) or any other superpotential that agrees with it to linear order.}
But \emph{none} of them gives positive energy for small vacuum regions~\cite{So:2009wu}.
An 11-parameter set of new pseudotensor superpotentials  with this desirable property was constructed by So~\cite{SoN09a}.

A widely held opinion was expressed in an influential textbook (Ref.~\cite{MTW}, p~467):
\textit{Anyone who looks for a magic formula for ``local gravitational energy-momentum'' is looking for the right answer to the wrong question.  Unhappily, enormous time and effort were devoted in the past to trying to ``answer this question'' before investigators realized the futility of the enterprise.}
Here we will present a somewhat different view.

We note
 that some time ago it was,
  \emph{surprisingly},
   found that the fundamentally different pseudotensors of Einstein, Landau-Lifshitz, Papapetrou and Weinberg could sometimes---\emph{in particular for all Kerr-Schild metrics}---give \emph{the same energy value}~\cite{Aguirregabiria:1995qz}.%
\footnote{Following the appearance of this seminal work, numerous papers have been continually appearing comparing the energy-momentum values given by various pseudotensors in various metrics.}
Our discovery is a far more general result---both with respect to the class of metrics and the class of expressions.

\section{The Hamiltonian approach}

How can one understand the physical significance of all these expressions?
The Hamiltonian approach offers a way.
Note that pseudotensors are related to the Hamiltonian.
This can be seen as follows.  For any region covered by a single coordinate system one can choose a vector field $Z^\nu$ with constant components in that frame.  The associated total energy-momentum, $P_\nu(V)$, in the region can then be
determined, using~(\ref{superpot}), to be
\begin{eqnarray}
- Z^\nu P_\nu(V) &:=& - \int_V Z^\nu ({{\mathfrak{T}}^\mu{}_\nu + {\mathfrak{t}}^\mu{}_\nu}) d\Sigma_\mu
\nonumber\\
&\equiv& \int_V Z^\nu |g|^{\frac12} \Bigl( \frac1{\kappa} G^\mu{}_\nu - T^\mu{}_\nu \Bigr) d\Sigma_\mu - \oint_{\partial V} Z^\nu {\mathfrak{U}^{\mu\lambda}{}_\nu}{\textstyle\frac12}dS_{\mu\lambda}
\nonumber\\
&\equiv& \int_V Z^\nu {\cal H}^{\rm GR}_\nu + \oint_{S=\partial V} {\cal B}(Z) \equiv H(Z, V). \label{basicHam}
\end{eqnarray}
Note that ${\cal H}^{\rm GR}_\nu$ is just the covariant expression for the ADM Hamiltonian density (see, e.g.~\cite{MTW} Ch.~21),
and the {\em boundary term} 2-surface integrand, ${\cal B}(Z)$,
is determined by the superpotential.
The value of the
pseudotensor/Hamiltonian is thus \textit{quasi-local}, determined just by this boundary term, since
the spatial volume integral vanishes ``on shell'' (initial value constraints).

From the Hamiltonian variation one gets information that tames the freedom in the boundary term---namely boundary conditions.
(The boundary term in the variation of the Hamiltonian indicates what must be held fixed on the boundary.)
The pseudotensor values thus are the values of the Hamiltonian with the associated boundary conditions~\cite{Chang:1998wj}. Hence the first problem  is under control: the generally different energy-momentum values from different pseudotensors correspond to the value of the associated Hamiltonians which evolve the system with their
respective boundary conditions.

\section{Quasi-local energy-momentum}

The modern concept is \emph{quasi-local} energy-momentum: i.e., associated with a closed 2-surface (pseudotensors always had this property, but this was not easily recognized before Freud's paper~\cite{Freud,Goldberg:1958zz}, and its importance only became appreciated after Penrose~\cite{Penrose}, where the term \emph{quasi-local} was introduced).
A comprehensive review~\cite{Sza09} states that we
\textit{``have
no ultimate, generally accepted expression for the energy-momentum and especially for the angular
momentum, ...''}
Some of the frequently proposed criteria for conserved quasi-local energy-momentum expressions will play a role in our presentation:
(i) they should vanish for Minkowski, (ii) give the standard linearized theory values at spatial infinity, and (iii) positive energy.

Notable quasi-local energy expressions include Komar~\cite{Komar},
M{\o}ller's tetrad-teleparallel expression~\cite{Mol61},
the spinor Hamiltonian boundary term 2-form associated with the Witten positive energy proof~\cite{spinorham},
the teleparallel gauge current~~\cite{duan,Wallner,deAndrade:2000kr},
Brown and York~\cite{BrownYork},
Bi\v{c}\'ak, Katz and Lynden-Bell~\cite{LBKB95} (equivalent to our favored expression discussed below), our 2-parameter set~\cite{Chen:1994qg,Chen:1998aw,Chen:2005hwa,GR100},
Tung's expression from his quadratic spinor Lagangian~\cite{Nester:1994zn},
Kijowski's ``free energy''~\cite{Kijowski97},
Epp's energy~\cite{Epp:2000zr},
the ``new superpotential'' of Petrov-Katz~\cite{Petrov:1999va},
the (\textit{positive}) energy of Liu-Yau~\cite{LiuYau} (the same as Kijowski's free energy),
and the Wang-Yau mass~\cite{WY09}.

\section{The covariant Hamiltonian formulation}
To provide a framework for our discovery
we briefly review some parts of our covariant Hamiltonian formulation~\cite{Chang:1998wj,Chen:1994qg,Chen:1998aw,Chen:2005hwa,GR100}.
From a first order Lagrangian ${\cal L} = {\rm d}\varphi \wedge p - \Lambda(\varphi, p)$ for a $k$-form field $\varphi$, by variation one obtains a pair of first order dynamical equations:
\begin{eqnarray}
\delta{\cal L} &=& {\rm d}(\delta\varphi\wedge p) + \delta\varphi \wedge \frac{\delta{\cal L}}{\delta\varphi} + \frac{\delta{\cal L}}{\delta p} \wedge \delta p. \label{deltaL}
\end{eqnarray}
From an infinitesimal diffeomorphism, a ``local translation'' along a vector field $Z$, with the aid of the formula for the Lie derivative acting on forms, $\pounds_Z \equiv i_Z {\rm d} + {\rm d} i_Z$, one obtains an identity by replacing $\delta$ in (\ref{deltaL}) with $\pounds_Z$:
\begin{equation}
{\rm d} i_Z {\cal L} \equiv \pounds_Z {\cal L} \equiv {\rm d}(\pounds_Z \varphi \wedge p) + \pounds_Z \varphi \wedge \frac{\delta{\cal L}}{\delta\varphi} + \frac{\delta{\cal L}}{\delta p} \wedge \pounds_Z p.
\end{equation}
This leads to the identification of the Noether current 3-form, which, moreover, is the Hamiltonian density:
\begin{equation}
{\cal H}[Z] := \pounds_Z \varphi \wedge p - i_Z {\cal L}. \label{5:HN}
\end{equation}
It satisfies the differential identity
\begin{equation}
- {\rm d} {\cal H}[Z] \equiv \pounds_Z \varphi \wedge \frac{\delta {\cal L}}{\delta \varphi} + \frac{\delta {\cal L}}{\delta p} \wedge \pounds_Z p, \label{dHZ}
\end{equation}
so it is a conserved ``current'' {\it on shell} (i.e., when the field equations are satisfied).
This $3$-form is linear in the displacement vector field up to a total
differential:\footnote{Where $\zeta := (-1)^k$.}
\begin{equation}
{\cal H}[Z] \equiv \zeta i_Z \varphi \wedge {\rm d}p + \zeta {\rm d}\varphi \wedge i_Z p + i_Z \Lambda + {\rm d}(i_Z\varphi \wedge p) =: Z^\mu {\cal H}_\mu + {\rm d} {\cal B}[Z]. \label{5:H+dB}
\end{equation}
Comparing the differential of the latter form, ${\rm d}{\cal H}[Z] \equiv {\rm d}Z^\mu \wedge {\cal H}_\mu + Z^\mu {\rm d}{\cal H}_\mu$, with (\ref{dHZ}) shows that
from {\em local\/} diffeomorphism invariance ${\cal H}_\mu$ is proportional to field equations and thus vanishes \emph{on shell}; hence the translational Noether current conservation reduces to a differential identity between Euler-Lagrange expressions (an instance of Noether's second theorem).
The value of the Hamiltonian $H(Z,V)$, which determines the energy-momentum, is thus \emph{quasi-local} (associated with a closed 2-surface):
\begin{equation}
-P(Z, V) = H(Z,V) := \int_V {\cal H}[Z] = \oint_{\partial V} {\cal B}[Z]. \label{5:EN}
\end{equation}

As with other Noether conserved currents, without loss of the conservation property one can add a ``curl'' to ${\cal H}(Z)$, i.e., the boundary term inherited from the Lagrangian, $i_Z \varphi \wedge p$, can be adjusted---this changes the boundary conditions that are obtained from the boundary term in the variation of the Hamiltonian.

That $H(Z,V)$, the integral of ${\cal H}[Z]$, is indeed the Hamiltonian follows from the easily established identity~\cite{Chen:2005hwa,GR100}:
\begin{equation}
\delta{\cal H}[Z] = - \delta\varphi \wedge \pounds_Z p + \pounds_Z \varphi \wedge \delta p + {\rm d} i_Z(\delta\varphi \wedge p) - i_Z \Bigl(\delta\varphi \wedge \frac{{\cal L}}{\delta\varphi} + \frac{{\cal L}}{\delta p} \wedge \delta p \Bigr).
\end{equation}
On shell, the total differential term gives rise to a boundary term that vanishes for the boundary condition $\delta\varphi|_S=0$.  With a modified boundary term, ${\cal B}'[Z]$, the Hamiltonian will yield a modified boundary condition.

We were led to a set of general boundary terms which are linear in $\Delta\varphi := \varphi - \bar\varphi$, $\Delta p := p - \bar p$, where $\bar\varphi, \bar p$ are non-dynamic reference values.
For Einstein's GR our ``covariant-symplectic'' Hamiltonian boundary terms are
\begin{equation} \label{Bcosymp}
2 \kappa {\cal B}[Z](a,b) = \Delta\Gamma^{\alpha}{}_{\beta} \wedge i_Z [ (1-a) \eta_{\alpha}{}^{\beta} + a \bar\eta_{\alpha}{}^{\beta} ] + [ (1-b) {\bar \nabla_{\beta} Z^\alpha} + b {\nabla_{\beta} Z^\alpha} ] \Delta\eta_{\alpha}{}^\beta,
\end{equation}
where $\eta^{\alpha\beta\dots} := *(\vartheta^\alpha \wedge \vartheta^\beta\dots)$, $\vartheta^\mu$ is the coframe and $\Gamma^\alpha{}_{\beta}$ is the connection one-form.
Here one may freely choose $a,b$. The choices $(0,0), (0,1), (1,0), (1,1)$ select
essentially Dirichlet (fixed field) or Neumann (fixed momentum) boundary conditions for the space and time parts of the fields~\cite{So:2006zz}.
For asymptotically flat spaces the Hamiltonian with these boundary term expressions is \emph{well defined}, i.e., the boundary term in its variation vanishes and the quasi-local quantities are well defined---at least on the phase space of fields satisfying Regge-Teitelboim~\cite{Regge:1974zd} like asymptotic parity and fall-off conditions.

Our preferred GR Hamiltonian boundary term is
\begin{equation}
2\kappa{\cal B}[Z] = \Delta\Gamma^{\alpha}{}_{\beta} \wedge i_Z \eta_{\alpha}{}^{\beta} + \bar \nabla_{\beta} Z^\alpha \Delta\eta_{\alpha}{}^\beta. \label{BprefGR}
\end{equation}
It corresponds to holding the metric fixed on the boundary.
Like many other boundary term choices, at spatial infinity it gives the ADM, MTW~\cite{MTW} \S 20.2, Regge-Teitelboim~\cite{Regge:1974zd}, Beig-\'O~Murchadha~\cite{Beig:1987zz}, Szabados~\cite{Szabados:2003yn} energy, momentum, angular-momentum, center-of-mass.
One of its special virtues is that at null infinity it gives the Bondi-Trautman energy and energy flux~\cite{Chen:2005hwa}.

\section{The reference}
Regarding the \emph{second ambiguity} inherent in the discussed quasi-local energy-momentum expressions, \emph{the choice of reference}:  one could use any physically appropriate reference, preferably a very symmetrical one.
If the chosen reference is a space of \emph{constant curvature} (de Sitter, anti-de Sitter or Minkowski) one has 10 reference Killing vector fields that can be used for the vector $Z$ to define all 10 quasi-local quantities: energy-momentum, angular momentum/center-of-mass for a Minkowski reference.

Here we consider only a Minkowski reference.
But how can one fix which Minkowski space?  Recently we proposed
(i) 4D isometric matching on the boundary and (ii) energy optimization
as criteria for the ``best matched'' reference on the region's boundary~\cite{Nester:2012zi,Sun:2015mxg}.

In a neighborhood of the desired spacelike boundary 2-surface $S$, any 4 smooth functions $ y^{\bar \mu} = y^{\bar\mu}(x^\nu), \; \bar\mu = 0, 1, 2, 3$ with ${\rm d}y^0 \wedge {\rm d}y^1 \wedge {\rm d}y^2 \wedge {\rm d}y^3 \ne 0$ define a Minkowski reference
$ \bar g = -({\rm d}y^0)^2 + ({\rm d}y^1)^2 + ({\rm d}y^2)^2 + ({\rm d}y^3)^2 $.
Locally this defines an embedding of a neighborhood of $S$ into a Minkowski space.
A Killing field of $\bar g$ has the infinitesimal Poincar\'e transformation form:
$\bar Z^{\bar \mu} = \alpha^{\bar\mu} + \lambda^{\bar\mu}{}_{\bar\nu} y^{\bar\nu}$,
where $\alpha^{\bar\mu}$ and $\lambda_{\bar\mu\bar\nu} \equiv \lambda_{[\bar\mu\bar\nu]}$ are constants.
Then the Hamiltonian boundary term quasi-local quantity integral has the form
\begin{equation}
E(\bar Z,S) = \oint_S {\cal B}[\bar Z] = - \alpha^{\bar\mu} p_{\bar\mu}(S) + \frac12 \lambda_{\bar\mu\bar\nu} J^{\bar\mu\bar\nu}(S), \label{QQ}
\end{equation}
giving the quasi-local energy-momentum and angular momentum values associated with this reference.  With suitable fall-offs, the integrals $p_{\bar\mu}(S)$, $J^{\bar\mu\bar\nu}(S)$ for our expressions agree with the standard expressions asymptotically~\cite{MTW,Regge:1974zd,Beig:1987zz,Szabados:2003yn}.

The reference metric on the dynamical space has the components
$ \bar g_{\mu\nu} = \bar g_{\bar\alpha\bar\beta} y^{\bar\alpha}_{~\mu} y^{\bar\beta}_{~\nu} $ and
the reference connection one-form (with ${\rm d}x^\mu = x^\mu_{~\bar\alpha} {\rm d}y^{\bar\alpha}$, ${\rm d}y^{\bar\alpha} = y^{\bar\alpha}_{~\nu} {\rm d}x^\nu$)
is
\begin{equation}
\bar \Gamma^\alpha{}_{\beta} = x^\alpha{}_{\bar\mu} ( \bar\Gamma^{\bar\mu}{}_{\bar\nu} y^{\bar\nu}{}_{\beta} + {\rm d}y^{\bar\mu}{}_{\beta} ) = x^\alpha{}_{\bar\mu} {\rm d}y^{\bar\mu}{}_{\beta},
\end{equation}
since the Minkowski connection coefficients $\bar\Gamma^{\bar\mu}{}_{\bar\nu}$ vanish in the $y^{\bar\mu}$ frame.
With $\bar Z^\mu$ being a translational Killing field of the Minkowski reference, the second term in~(\ref{BprefGR}) vanishes, then
our quasi-local expression takes the form
\begin{eqnarray}
2\kappa{\cal B}(\bar Z) &=& \bar Z^{\bar\nu} x^\mu{}_{\bar\nu} (\Gamma^\alpha{}_{\beta} - x^\alpha{}_{\bar\rho} \, {\rm d}y^{\bar\rho}{}_{\beta}) \wedge \eta_{\mu\alpha}{}^\beta \label{BMink}
\\
&\equiv& \bar Z^{\bar\nu} \Gamma^{\bar\alpha}{}_{\bar\rho} \wedge \eta_{\bar\nu\bar\alpha}{}^{\bar\rho}
\equiv \bar Z^{\bar\nu} |g|^{\frac12} g^{\bar\rho\bar\sigma} \Gamma^{\bar\alpha}{}_{\bar\rho \bar\gamma} \delta^{\bar\mu \bar\kappa \bar\gamma}_{\bar\alpha \bar\sigma \bar\nu} {\textstyle\frac12}dS_{\bar\mu \bar\kappa}. \label{BF}
\end{eqnarray}
Thus, when expressed in the Minkowski reference coordinate frame, it reduces to the Freud superpotential~(\ref{UF}).

Our first criterion for fixing the reference is 4D isometric matching to a Minkowski reference on the boundary of the region.
The hard part is the isometric embedding of the 2D surface $S$ into Minkowski space (Wang and Yau have made deep investigations into this~\cite{WY09}).
For isometric matching of the 2-surface, in terms of quasi-spherical foliation adapted coordinates $t, r, \theta, \phi$ with $i,j = 1,2,3$ and $A,B = \theta,\phi$:
\begin{equation}
g_{AB} = \bar g_{AB} = \bar g_{\bar\mu\bar\nu} y^{\bar\mu}{}_{A} y^{\bar\nu}{}_{B} = - y^0{}_{A} y^0{}_{B} + \delta_{ij} y^i{}_{A} y^j{}_{B}
\end{equation}
on $S$.
From a classic closed 2-surface into $\mathbb R^3$ embedding theorem, there is a unique embedding (\emph{but no explicit formula}) ---as long as the choice of $S$ and $y^0$ are such that
$ g_{AB}' := g_{AB} + y^0{}_{A} y^0{}_{B} $
is convex on $S$.  If this is not satisfied the isometric embedding is not guaranteed.

Complete 4D isometric matching on $S$ was proposed in 2000 by Epp~\cite{Epp:2000zr}
and by Szabados.\footnote{Szabados,~L.B.: in a talk at Tsinghua Univ., Hsinchu, Taiwan, July 2000.;
  Szabados,~L.B.: ``Quasi-local energy-momentum and angular momentum in GR: the covariant Lagrangian approach,''
  unpublished draft, 2005.}
There are 10 constraints:
$ g_{\alpha\beta}|_S = \bar g_{\alpha\beta}|_S = \bar g_{\bar\mu\bar\nu} y^{\bar\mu}{}_\alpha y^{\bar\nu}{}_\beta|_S $
and 12 embedding functions  on a constant $t,r$ 2-surface:
$ y^{\bar\mu} (\Longrightarrow y^{\bar\mu}{}_{\theta}, y^{\bar\mu}{}_{\phi}), \ y^{\bar\mu}{}_{t}, \ y^{\bar\mu}{}_{r}$.
The 10 constraints split into 3 for the already discussed 2D isometric matching of $g_{AB}$
whereby $y^0$ determines $y^1, y^2, y^3$ on $S$,
and 7 algebraic equations that determine the other embedding variables. 
One can take $y^0, y^0{}_r$ as the embedding control variables; $y^0{}_r$ controls a boost in the normal plane~\cite{Nester:2012zi,Sun:2015mxg}.

An alternative regards 4D isometric matching in terms of orthonormal frames.  The reference geometry has a frame of the form $\bar\vartheta^{\bar\mu} = {\rm d}y^{\bar\mu}$.  With 4D isometric matching one can choose a dynamical $\vartheta^{\hat\alpha}$ frame that can be Lorentz transformed to match such a reference frame on $S$:
\begin{equation}
L^{\bar\mu}{}_{\hat\alpha}(p) \vartheta^{\hat\alpha}(p) = \bar\vartheta^{\bar\mu}(p) = {\rm d}y^{\bar\mu}(p), \qquad \forall p \in S.
\end{equation}
Then restricted to $S$ one has four integrability conditions:
\begin{equation}
{\rm d}(L^{\bar\mu}{}_{\hat\alpha} \vartheta^{\hat\alpha})|_S = 0,
\end{equation}
each a one-component 2-form, thus 4 restrictions on 6 Lorentz parameters, hence again 2 degrees of freedom.

\section{A common value}

We have reviewed some of the history and surveyed many of the proposed energy expressions and have laid out the Hamiltonian perspective as a framework for clarifying the issues. With this foundation we can now easily explain our remarkable discovery.

\emph{With a 4D isometric matching reference  the values of many distinct expressions coincide}. This is obvious for the covariant symplectic Hamiltonian boundary terms~(\ref{Bcosymp}),
as one then has $\eta_\alpha{}^\beta|_S=\bar\eta_\alpha{}^\beta|_S$.
If the reference is Minkowski the expression~(\ref{BprefGR}) becomes~(\ref{BF}), which is holonomically the Freud expression~(\ref{UF}, \ref{UF3}), and orthonormally it is the teleparallel gauge current. Furthermore,
with $g_{\mu\nu}|_S=\bar g_{\mu\nu}|_S$
the superpotentials (\ref{LL}), (\ref{HP}), (\ref{UBT}), (\ref{UFn}), (\ref{HW}) will then coincide on $S$.  Checking other expressions we discovered a surprising concord.

\emph{For any closed 2-surface in a dynamical Riemannian spacetime, with 4D isometric matching to a Minkowski reference, there is a
common quasi-local energy value for all the expressions that linearly agree with the Freud superpotential, (\ref{UF}) or (\ref{UF3}), in the Minkowski limit}.

Linearly  Freud-like  expressions having a concordant value include:
 Landau-Lifshitz~(\ref{LL}),
  Papapetrou~(\ref{HP}),
 Bergmann-Thomson~(\ref{UBT}),
Goldberg's weighted Freud densities~(\ref{UFn}),
M{\o}ller's 1961~\cite{Mol61} tetrad-teleparallel expression,
\begin{equation}
{\cal B}_{\rm M61}(\partial_\nu) = \Gamma^{\hat\alpha}{}_{\hat\beta} \wedge e^{\hat\mu}{}_\nu \eta_{\hat\mu\hat\alpha}{}^{\hat\beta},
\end{equation}
the {teleparallel gauge current}~\cite{duan,Wallner,deAndrade:2000kr},
\begin{equation}
{\cal B}_{\|}(e_{\hat\mu}) = \Gamma^{\hat\alpha}{}_{\hat\beta} \wedge \eta_{\hat\mu\hat\alpha}{}^{\hat\beta},
\end{equation}
Weinberg~(\ref{HW}),
the spinor Hamiltonian boundary term 2-form associated with the Witten positive energy proof~\cite{spinorham}:\footnote{Here $\gamma_{0123} := \gamma_0 \gamma_1 \gamma_2 \gamma_3$.}
\begin{eqnarray}
{\cal B}_\psi &=& 2 [ \bar\psi \gamma_{0123} \gamma_{\hat\mu} \vartheta^{\hat\mu} \wedge D\psi + D\bar\psi \wedge \gamma_{0123} \gamma_{\hat\mu} \vartheta^{\hat\mu} \psi ]
\nonumber\\
&\equiv& 2 [ \bar\psi \gamma_{0123} \gamma_{\hat\mu} \vartheta^{\hat\mu} \wedge {\rm d}\psi + {\rm d}\bar\psi \wedge \gamma_{0123} \gamma_{\hat\mu} \vartheta^{\hat\mu} \psi ] + \bar\psi \gamma^{\hat\mu} \psi \Gamma^{\hat\alpha\hat\beta} \wedge \eta_{\hat\alpha\hat\beta\hat\mu}, \label{BSW}
\end{eqnarray}
for a spinor having \emph{constant components} in the Minkowsi reference frame on the boundary,
Tung's spinor expression~\cite{Nester:1994zn},
our 2-parameter covariant-symplectic boundary expressions~(\ref{Bcosymp}),
Bi{\v{c}}\'ak-Katz-Lynden-Bell~\cite{LBKB95},
Petrov-Katz~\cite{Petrov:1999va} (essentially it replaces the first term in (\ref{BprefGR}) by (\ref{HP}), the Papapetrou superpotential),
So's 11-parameter superpotentials~\cite{SoN09a}, and, as we shall explain,  Wang-Yau~\cite{WY09}.

In view of the quasi-local desiderata of (i) vanishing for Minkowski and (ii) agreeing with the standard spatial infinity linear results,
this concord could have been expected,
even though these various expressions differ beyond the linear order and may not agree for other than an isometric Minkowski reference. \emph{Different energy expressions can have a common value far more generally than Ref.~\cite{Aguirregabiria:1995qz} had imagined,
in terms of both the class of metrics and the class of expressions}.

Some pseudotensor/Hamiltonian-boundary term expressions give other values for the quasi-local energy. This includes those that do not have the
aforementioned desirable
spatial asymptotic linearized theory limit
or do not  choose the reference by embedding into Minkowski space, e.g,
M{\o}ller~(\ref{UM}),
Komar, ${\cal B}(Z) = *{\rm d}( Z_\mu \vartheta^\mu)$,~\cite{Komar},
the expression~(\ref{BSW}) (with the spinor boundary values needed for the Witten positivity proof),
Brown-York~\cite{BrownYork} with $S$ embedded into $\mathbb{R}^3$ reference,
Kijowski's ``free energy''\cite{Kijowski97}, Epp~\cite{Epp:2000zr}, and Liu-Yau~\cite{LiuYau}.
Also, we should mention that some well-known quasi-local proposals are not formulated in the Lagrangian/Hamiltonian framework, e.g.
Penrose's twistor expression~\cite{Penrose},
Hawking's mass~\cite{Hawking}, and
Hayward's expression~\cite{Hayward}.

\section{A best matched reference}

What has been described in the previous section is not a specific reference, but a whole class of references.
Within the set of isometric Minkowski references
can one find a ``best matched'' Minkowski reference geometry?
For any of the common value expressions, e.g.~(\ref{BMink}), there are   12 embedding variables subject to 10 isometric conditions (or 6 orthonormal frame parameters subject to 4 conditions). To determine the two embedding control variables,
one can use the boundary term value.
Its critical points are distinguished~\cite{Nester:2012zi,Sun:2015mxg} and can be used to select a specific reference.

There are 2 quantities which could be considered: $m^2 c^2 = - \bar g{}^{\bar\mu\bar\nu} p_{\bar\mu} p_{\bar\nu} = p_0^2 - p_1^2 - p_2^2 - p_3^2$ and $p_0$.
Technically $m^2 c^2$ is more complicated: one is extremizing a linear combination of quadratic quantities, each an integral over $S$;  this would determine the reference up to Poincar\'e transformations.  The Lorentz ``gauge'' freedom could then be used to specialize to the frame with vanishing momentum: $\vec p = 0$.  In this ``center-of-momentum'' frame $m^2c^2$ reduces to $p_0^2$.  But the critical points of $p_0$ are also  critical points of $p^2_0$.
So one can get \textit{the same reference} from the much more simple expression  $-c p_0 = E(\partial_{y^0}, S)$, which is a smooth function of the 2 embedding control variables.

From our 4D approach we have not found a general analytic formula for the critical points,
however for the special case of \emph{axisymmetric} metrics (including Kerr) we can explicitly find the critical point analytically~\cite{Sun:2013ika}. Do suitable critical points generally exist? Can one find them?
Here is a practical computational argument.
Consider being given a set of data from a numerical relativity calculation. Compute the energy given by a large number of reference choices; the critical value(s) will stand out.

\subsection{Wang-Yau}

From another perspective, in a milestone work
Wang and Yau~\cite{WY09}
used a quasi-local expression in terms of surface geometric quantities (metric, normals, extrinsic curvatures)
based on the Hamilton-Jacobi approach of Brown and York~\cite{BrownYork} as developed in~\cite{Hawking:1995fd}. They found a way to determine an optimally embedded isometric Minkowski reference analytically and thereby obtained their \emph{quasi-local mass};  moreover,  they were able to show that their quasi-local mass  is \textit{non-negative} and, furthermore, \textit{vanishes for Minkowski}.
An outstanding achievement.

\subsection{The link}

In a recent work Liu and Yu~\cite{Liu:2017neh} found that the expression~(\ref{BprefGR})
with a 4D isometric matching Minkowski reference
is closely related to the expression used by Wang and Yau;
consequently a saddle critical value of the associated energy
agrees with the Wang-Yau mass. This is not very surprising, as both approaches start with the Einstein-Hilbert Lagrangian of GR and follow (albeit different) paths that lead to a GR Hamiltonian boundary term. Nevertheless this is an important link, especially since, as we indicated, all the linearly-like-Freud expressions with a 4D isometric Minkowski reference give the same energy value as the expression~(\ref{BprefGR}).
We now realize that the results was not a special property of the expression~(\ref{BprefGR}); one could have used any other Freud linear expression and established the link had they thought to do so.

Hence there is,
thanks to~\cite{WY09} (with the link established by~\cite{Liu:2017neh})
\textit{for all the  Freud-linear expressions} a specific quasi-local energy, which
satisfies the main criteria:
\emph{it is non-negative and vanishes if the dynamical spacetime is Minkowski}.
This simple observation is an additional result on top of our earlier mentioned discovery.

\section{Concluding discussion}

We have surveyed much of the work that had been done regarding the unsettled issue of gravitational energy.
While the energy of gravitating systems cannot be localized, with the aid of the Hamiltonian framework we explained that with an appropriate Minkowski reference one can find a good quasi-local energy.

The quasi-local energy with nice properties referred to at the end of the previous section
could have been found even as early as 1939.  It was found only 70 years later by Wang and Yau~\cite{WY09} via a Hamilton-Jacobi approach. We came to this value a few years later by a 4D covariant Hamiltonian route~\cite{Nester:2012zi}.
Einstein with his pseudotensor was close (as were many others), merely lacking a good way to choose the coordinate reference frame on the boundary.

As noted, much effort by many people was expended in trying to find the ``best'' energy-momentum expression.
Our group investigated the roles of the Hamiltonian boundary term. We found a preferred 4D covariant (reference dependent) expression.
Only then did we turn to finding a good reference.
But just taking (almost) any of the proposed expressions and looking for the ``best'' reference could have led anyone directly to this energy.

The complaint was that gravitational energy is ill defined:
(i) no unique expression,
(ii) reference frame dependent expressions with no unique reference frame.
But we find that: (a) one can generally have a 4D isometric Minkowski reference,
(b) with such a reference the quasi-local expressions in a large class give the same energy-momentum (and via (\ref{QQ}) angular momentum), (c) and one can find a ``best matched'' reference.

As the textbook said: \emph{there is no proper local energy-momentum density}---now it is understood that energy-momentum is not local but rather \emph{quasi-local}---and, although much effort was expended on finding the best expression, indeed there still is no single accepted quasi-local expression.  However, we discovered that \emph{a consensus on the expression is not needed}---associated with any region for a large class\footnote{Much larger than had found previously in~\cite{Aguirregabiria:1995qz}.}
 of reasonable proposed expressions (those which have the desired asymptotics) \emph{there is a well defined energy value} (fixed by the 4D best-matched isometric Minkowski reference), which does have the desired properties.

\section*{Acknowledgements}

C.M.C. was supported by the Ministry of Science and Technology of the R.O.C.
under the grant MOST 106-2112-M-008-010. J.-L. Liu was supported by the China Postdoctoral Science Foundation 2016M602497 and partially supported by the National Natural Science Foundation of China 61601275.

\end{document}